**Identifying RNA contacts from SHAPE-MaP by partial correlation analysis**

Akshay Tambe[a], Jennifer Doudna[a,b] and Lior Pachter[a,c*]

a: Department of Molecular & Cell Biology, UC Berkeley.
b: Department of Chemistry, UC Berkeley.
c: Department of Mathematics and Computer Science, UC Berkeley.
*: To whom correspondence should be addressed (lpachter@math.berkeley.edu).

In a recent paper Siegfried *et al.*[1] published a new sequence-based structural RNA assay that utilizes mutational profiling to detect base pairing (MaP). Output from MaP provides information about both pairing (via reactivities) and contact (via correlations). Reactivities can be coupled to partition function folding models for structural inference[2], while correlations can reveal pairs of sites that may be in structural proximity[3]. The possibility for inference of 3D contacts via MaP suggests a novel approach to structural prediction for RNA analogous to covariance structural prediction for proteins[4].

To assess whether SHAPE-MaP data contain pairwise 3D structural information, we first measured the area under the precision-recall curves (AUPR) using site correlations and the corresponding contact maps for the crystallized molecules in the Siegfried *et al.* and Homan *et al.*[3] datasets. The AUPR values at 15 Å for the Siegfried *et al.* data are 0.3 for the 16S rRNA, 0.67 for the 23S rRNA and 0.62 for the 5S rRNA. In all cases the numbers reflect the fact that the SHAPE-MaP correlations and contacts are clustered at sites that are neighbors along the respective molecules. The correlation based AUPR values are smaller for the Homan *et al.* data (Figure 1a) but the correlation matrices and crystallographic structures indicate that non-linear contacts could in principle be determined.

While it is intuitive to associate highly correlated MaP sites with structural contacts, it is well known[5] that the precision matrix is better suited to inferring interacting sites than the correlation matrix. We therefore asked whether partial correlation analysis could improve on correlation analysis for inference of contact maps. Since the molecules in Siegfried *et al.* were not suitable for such an analysis we focused on the Homan *et al.* data. Figure 1a shows improvement in AUPR for all three molecules in the paper, with partial correlation applied to RNaseP showing significant improvement in matching the crystal contact map (Figure 1b). While correlation analysis identifies only 26% of true contacts at a false positive rate of 20%, partial correlation analysis can identify 52% of true contacts with the same false positive rate.

The overall improvement of partial correlation analysis over correlation analysis is consistent with results from protein structural prediction and gene regulatory network analysis. Even though the most highly correlated sites are in contact, some correlations may result from transitive effects: when A is correlated with B and B is correlated with C then A is correlated with C. This is evident in Figure 3c of Homan *et al.* via the presence of many triangles. As shown by example (Figure 1c, d), partial correlations greatly reduce this effect. This suggests that partial correlation analysis may become a powerful tool for structural inference using SHAPE-MaP.

1. Siegfried, N. A., Busan, S., Rice, G. M., Nelson, J. A. E. & Weeks, K. M. *Nat Meth* **11,** 959–965 (2014).

2. Hajdin, C. E. *et al*. *PNAS* **110,** 5498–5503 (2013).

3. Homan, P. J. *et al.* *PNAS* **111**, 13858–13863 (2014).

4. Göbel, U., Sander, C., Schneider, R. & Valencia, A. *Proteins* **18,** 309–317 (1994).

5. Lauritzen, S. L. (1996).

6. Schäfer, J. & Strimmer, K. *Bioinformatics* **21,** 754–764 (2005).

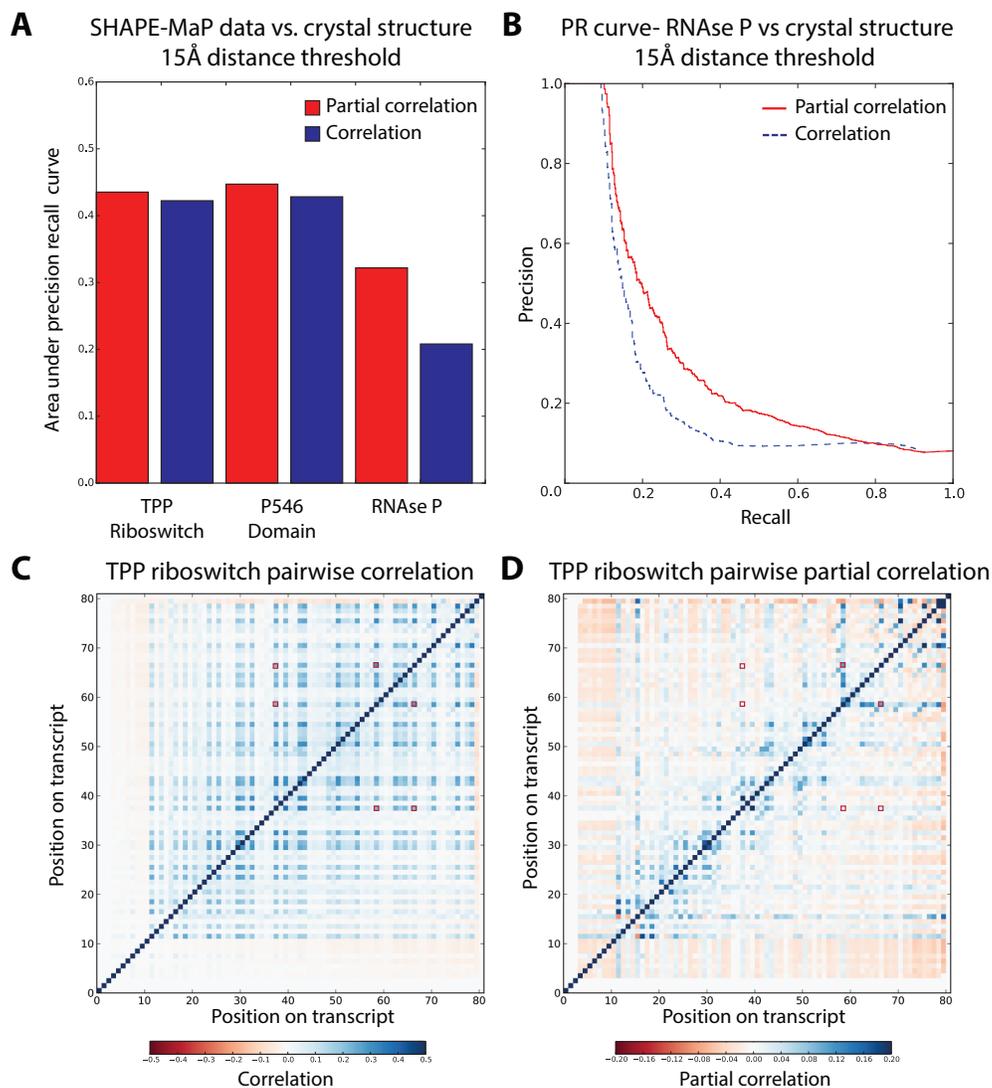

Figure 1: a) Area under the precision-recall curve (AUPR) for correlation and partial correlation analysis. AUPR was calculated using all base pairs within 15 Å for each of the molecules. b) The precision-recall curves for the RNaseP molecule showing significant improvement of partial correlation versus correlation. c) The correlation matrix for the TPP riboswitch showing the difficulty in resolving real contacts due to the transitivity of correlation. The red boxes mark three contact edges shown as a triangle in Figure 3c of Homan *et al.* d) The partial correlation matrix for the TPP riboswitch with the same red boxes showing how edges in triangles are "cleaned up" to reveal a more accurate contact map.